\documentstyle[12pt]{article}
\begin{document}
\thispagestyle{empty}
\begin{center}
\LARGE \tt \bf {Photon mass new limits from strong photon-torsion coupling generation of primordial magnetic fields}
\end{center}

\vspace{1.0cm}

\begin{center}
{\large By L.C. Garcia de Andrade\footnote{Departamento de F\'{\i}sica Te\'{o}rica - IF - UERJ - Rua S\~{a}o Francisco Xavier 524, Rio de Janeiro, RJ, Maracan\~{a}, CEP:20550.e-mail:garcia@dft.if.uerj.br}}
\end{center}

\begin{abstract}
 Recently Adelberger et al [Phys Rev Lett 98: 010402, (2007)] have placed a limit to photon mass by investigating the primordial magnetic fields. Earlier Bertolami et al [Phys Lett \textbf{B} 455, 96(1999)] showed that massive photons in a spontaneous Lorentz breaking may generate primordial magnetic fields consistent with galactic dynamo seeds. Torsion coupling constant of order $10^{-5}$, much higher than the previously obtained by de Sabbata and Sivaram of $10^{-24}$, leads to strong amplification of magnetic field able to seed galactic dynamo at recombination era contrary to what happens in general relativistic dynamos. This results in $B\sim{10^{-5}{\beta}G}$ where ${\beta}$ is the massive photon-torsion coupling. Thus in order to obtain the observed galaxy field of $B_{G}\sim{{\mu}G}$ one should have a coupling $\beta\sim{10^{-1}}$, never observed in the universe. Thus we may conclude that the weaker couplings for torsion to e.m fields shall only produce magnetic fields without dynamos starting from extremely strong magnetic seeds. From the strongest photon-torsion considered one obtains the best CMB estimate for torsion generated magnetic fields $\frac{{\delta}B}{B}\le{10^{-4}}$. By making use of the strong photon-torsion limits obtained here, photon mass limit of $m_{\gamma}\sim{10^{-24}eV}$, well-within limits found in literature, which allows us to conclude that a stronger massive photon-torsion limit is physically consistent. Actually this last limit is also graviton mass limit. This results differs from Adelberger et al by two orders of magnitude.
\end{abstract}
\newpage

\section{Introduction}
In the seminal paper by Turner and Widrow \cite{1} several non-minimal coupling Lagrangeans between electromagnetic field and Riemannian curvature of Einstein's general relativity, have been proposed with the purpose of computing amplification of the magnetic field able to seed galactic dynamos \cite{2}. One of the most interesting of their models was the one where the massive photon breaks the gauge invariance. Inflationary models were used in their computations. To solve several drawbacks in their models, more recently Bassett et al \cite{3} proposed to investigate a preheating phase of the universe, where magnetic field amplification could lead to a cosmic magnetic dynamo. In their models of geometric magnetisation \cite{4} again a tiny photon mass was present. In this brief report following previous papers \cite{5} we try to place limits on a photon-torsion Lorentz violation constrained by the observational fact that a galactic dynamo seed would lead to galactic magnetic fields of the order of ${\mu}G$. More recently Fenu et al \cite{6} has shown that $10^{-29}G$ obtained at cluster scales at recombination era, is too weak to be amplified by a dynamo mechanism. Here following previous idea by Bertolami et al \cite{7} on the spontaneous Lorentz breaking with massive photons we compute a most stringent bound on massive photon-torsion coupling which yields the possibility of a strong torsion field at recombination era which in turn generates dynamo amplification of seed field which are extremely weak. Previously massless photons from radiogalactic sources seems also to be able to generate LV \cite{8}. In this paper by analogy with Jimenez et al \cite{9} we show that magnetic fields can be obtained by massive photon-torsion strong non-minimal coupling constant Lagrangean. Simple derivation of dynamo equation to allow for  tiny photon mass is given. In section II expression for the dynamo equation is derived. In section III diffusion time decay of the magnetic field is computed and the values massive photon-torsion coupling would have to assume in order that the magnetic field does not decay which is the case of marginal dynamo, and the value at recombination are computed. Photon mass limits are computed in this section as well. Section IV contains conclusions and discussions.
\newpage
\section{Non-minimal massive photon coupling to spacetime torsion and dynamo equation}
One of the most interesting links between torsion and high energy physics is via Kalb-Rammond fields in string theory \cite{10} nevertheless, this link which can even couple torsion eventually to string dynamos \cite{11}, leaves us with few contacts with experiments. To remedy this situation one in general lay hands on Einstein-Cartan theory where torsion in general does not propagate or to quadractic curvature lagrangeans where torsion does propagate. Here we take into account second order effects on torsion and assume torsion is inhomogeneous in the universe, which allows us to simplify the problem. Let us consider the Lagrangean where non-minimal photon-torsion coupling constant shall be determined a posteriori by assuming that the magnetic field seeds are strongly amplified by torsion fields at recombination in order to seed galactic dynamos, and that the decaying diffusion time of the magnetic field is around $10^{9}yr$ \cite{2}. The Lagrangean is given by
\begin{equation}
{\cal{S}}= \int{d^{4}x(-g)^{\frac{1}{2}}[(-\frac{1}{4}F^{\mu\nu}F_{\mu\nu}+{\beta}RA_{\mu}A^{\nu})+J^{\mu}A_{\mu}]}
\label{1}
\end{equation}
Here external current $J^{\mu}$ where $(\mu=0,1,2,3)$, and the the second term the RHS of the Lagrangean, takes into account interaction between cosmic photon-torsion plasma and primordial magnetic field. The const ${\beta}$ is the photon-torsion constant and $R$ is the Ricci scalar where solely torsion terms are taken into account due to the Minkowski spacetime assumption. Let us now compute the Maxwell-like equation w.r.t the magnetic vector potential $A_{\mu}$ as
\begin{equation}
{\partial}_{\mu}F^{{\mu}{\nu}}=J^{\nu}\label{2}
\end{equation}
where expression for the electric-gravitational current now reads
\begin{equation}
{J^{\nu}}={\beta}RA^{\nu}-T_{\mu}F^{\mu\nu}\label{3}
\end{equation}
Note that since we are considering here from now one that torsion $T=constant$, which favor LV analysis, the Ricci-Cartan scalar R is given by $R\sim{T^{2}}$. The other set of e.m equations are
\begin{equation}
{\partial}_{[\mu}F_{\rho\sigma]}=0
\label{4}
\end{equation}
these new Maxwell's equations coupled with torsion yield
\begin{equation}
{\partial}_{\mu}F^{\mu\nu}-{\beta}RA^{\nu}+T_{\mu}F^{\mu\nu}=0
\label{5}
\end{equation}
Note that this tensorial equation can be splitted into two distinct equations for $\nu=0$ and $\nu=1$, given respectively by
\begin{equation}
{\partial}_{i}F^{i0}-{\beta}RA^{0}+T_{i}F^{i0}=0
\label{6}
\end{equation}
and
\begin{equation}
{\partial}_{0}F^{0i}+{\partial}_{k}F^{ki}-{\beta}RA^{i}+T_{j}F^{j}=0
\label{7}
\end{equation}
where $i,j,k=1,2,3$. By substituting the electric and magnetic vectors $E_{i}=F_{0i}$ and$B_{i}={\epsilon}_{ijk}F^{jk}$, and by making use of temporal gauge $A^{0}=0$, one obtains the 3D equations
\begin{equation}
{\nabla}.\textbf{E}+\textbf{T}.\textbf{E}=\rho
\label{8}
\end{equation}
\begin{equation}
{\partial}_{t}\textbf{E}+{\nabla}{\times}\textbf{B}-\textbf{T}\times\textbf{B}={\beta}R\textbf{A}
\label{9}
\end{equation}
The other set of Maxwell's equations, in 3-vector notation yields
\begin{equation}
{\partial}_{t}\textbf{B}={\nabla}{\times}\textbf{E}
\label{10}
\end{equation}
\begin{equation}
{\nabla}.\textbf{B}=0
\label{11}
\end{equation}
Here of course, $\rho$ is the electric charge density. By considering the electric vector current $\textbf{J}$ given by
\begin{equation}
\textbf{J}=\sigma[\textbf{E}+\textbf{v}\times\textbf{B}]
\label{12}
\end{equation}
yields
\begin{equation}
\nabla{\times}{\nabla}\times\textbf{B}={\nabla}\times\textbf{J}-{\nabla}\times[\textbf{T}\times\textbf{B}]
\label{13}
\end{equation}
where $\sigma$ is the electric conductivity, and its inverse $\eta$ is the cosmic plasma
resistivity or diffusion. Simple algebraic manipulations of these equations yields the MHD dynamo equation
\begin{equation}
{\partial}_{t}\textbf{B}+{\nabla}{\times}[\textbf{v}\times\textbf{B}]-\eta[{\Delta}\textbf{B}-
{\nabla}\times[\textbf{T}\times\textbf{B}]-{\beta}R\textbf{B}]=0
\label{14}
\end{equation}
This expression keeps some resemblance with the analogous general relativistic one in MHD derived by
Marklund and Clarkson \cite{12}. In their expression the expressions under the brackets involve the Ricci scalar of the Riemannian spacetime, while ours involve only the torsion vector. Let us now consider that torsion vector $\textbf{T}$ is parallel to the magnetic induction vector $\textbf{B}$, then from the dimensional analysis one may write
\begin{equation}
{\partial}_{t}{B}-\eta[{\Delta}{B}-{\beta}R{B}]=0
\label{15}
\end{equation}
as
\begin{equation}
{\partial}_{t}{B}-\eta[\frac{1}{L^{2}}-{\beta}R]{B}=0
\label{16}
\end{equation}
where L is the coherent length for magnetic fields. Slow dynamo equation (\ref{16}) is an eigenvalue expression for magnetic vector B.  Assuming that the torsion term dominates the coherent length term, one obtains
\begin{equation}
{\partial}_{t}{B}+\eta{\beta}R{B}=0
\label{17}
\end{equation}
Thus since diffusion $\eta$ is positive along with the coupling constant, then the only hope that the magnetic $B$ does not decay was for negative Ricci scalar be negative. But at least for the torsion axial-vector this is not the case, so actually magnetic field decays. The solution of dynamo equation
\begin{equation}
B=B_{0}exp[-\eta{\beta}T^{2}{\Delta}t]
\label{18}
\end{equation}
where $B_{0}$ is the seed field which may be amplified or decayed in the universe. This result is effective due to the assumption that torsion vector modulus is constant. In the next section two consequences of this dynamo equation are presented the CMB estimates for the Lorentz violation and the determination of $\beta$ in several cases of interest.
\section{Diffusion decay of magnetic fields, photon mass limits and CMB}
 Since the term $\eta{\beta}R$ in expression (\ref{18}) is quite small due to the fact that $R\sim{T^{2}}$ and axial-torsion estimated by Laemmerzahl \cite{13} is $T\sim{10^{-17}cm^{-1}}\sim{10^{-31}GeV}$, this decay can be expressed as
\begin{equation}
|B^{torsion}|=|{B}_{seed}|\eta{\beta}T^{2}{\Delta}t
\label{19}
\end{equation}
where torsion in the upscript means that we are only considering second order effects on the decaying of the magnetic field. Now to derive the coupling constant $\beta$, one imposes that the diffusion time ${\tau}_{d}={\frac{B}{\eta{\nabla}^{2}B}}\sim{\frac{L^{2}}{\eta{\omega}^{2}t^{2}}}$, which taking into account the galactic
discs with ${\eta}\approx{10^{26}cm^{2}s^{-1}}$, coherence length of $L\sim{3kpc}$ and vorticity
${\omega}\sim{10^{-15}s^{-1}}$, yields the decay time $t\sim{{\tau}_{d}}\sim{10^{8}yrs}$, which is much shorter than the
age of a galaxy. Now in torsion case analogously we may use the second term in the above dynamo equation inside the brackets to compute the diffusion scale in torsion case as
\begin{equation}
{\tau}_{d}=\frac{B}{\eta{{\nabla}^{2}}B}\sim{\frac{1}{\beta{\eta{T^{2}}}}}
\label{20}
\end{equation}
which by using the torsion value estimated by Laemmerzahl \cite{13}, given by $T\sim{10^{-31}GeV}$ allows to obtain for recombination time $10^{13}s$, an estimate for the massive photon-torsion coupling $\beta$. In order for example that the magnetic field does not decay the torsion coupling should be computed from expression (\ref{18}) making the $B_{seed}:=B^{torsion}$ which yields ${\beta}_{rec}\sim{10^{-5}}$. This coupling is much stronger than the coupling discovered by de Sabbata and Sivaram \cite{14} of $10^{-24}$. In what follows we shall show that by assuming that the massive photon-torsion coupling be effective in amplifying the magnetic fields in order that $B^{torsion}:=B_{G}\sim{10^{-6}G}$ one obtains ${\beta}_{postrec}\sim{10^{-1}}$ for the post-recombination where $B_{seed}\sim{10^{-8}G}$. Before to compute CMB from these two values, let see how this massive photon-torsion couplings affect the diffusion time. From expression (\ref{20}) one obtains ${\tau}_{d}\sim{10^{2}yrs}$ for post-recombination era which shows that the decay is too fast for $\beta\sim{10^{-5}}$ which is much less than the galaxy structure formation as above. A general expression for the magnetic field in terms of torsion can be obtained in recombination era as $B\sim{10^{-5}{\beta}G}$, which is similar to Jimenez-Maroto \cite{12} expression $B\sim{10^{-4}{\beta}G}$ using PPN approximation general relativistic method. In their case also the dynamo amplification is not needed, however, instead of getting magnetic field decay they obtained some magnetic field amplification. From expression
\begin{equation}
B=B_{0}[1-\eta{\beta}T^{2}{\Delta}t]
\label{21}
\end{equation}
which yields a CMB magnetic contrast \cite{15} as
\begin{equation}
|\frac{{\delta}B}{B}|=\eta{\beta}T^{2}{\Delta}t
\label{22}
\end{equation}
Note that the use of modulus is to better compare with COBE data, actually this does not mean that the contrast is positive which would be the presence of dynamo action. Substitution of the values of $\beta$ above reveals that the stronger the massive photon torsion coupling the more closest is the value of COBE above. Last but certainly not least is now the expression of photon mass
\begin{equation}
{m^{2}}_{\gamma}\sim{{\beta}T^{2}}
\label{23}
\end{equation}
From the limits discussed above and the photon-torsion coupling of $10^{-5}$ one easily obtains a photon mass limit of $10^{-24}eV$ which is well within limits found in the literature \cite{16} as $10^{-27}eV\le{m_{\gamma}}\le{10^{-13}eV}$. Of course limits found by using de Sabbata-Sivaram limit would left us with a limit $10^{-34}eV$ a limit never found in nature. Actually it is interesting to note that our limit coincides with the graviton mass limit \cite{16}. Adelberger et al \cite{17} have considered that the galactic and primordial magnetic fields vortices could destroy photon mass bounds and improve previous limits to $10^{-26}eV$ two orders of magnitude more stringent than ours. Nevertheless yet stronger photon-torsion couplings might bit this limit.
\section{Discussion and conclusions}
In this brief report we examined the issue of the improving massive photon-torsion coupling from old estimates by De Sabbata and Sivaram \cite{14} by astrophysical constraints. Actually from the above expressions is easy to show that by assuming that their estimate is right at recombination, a value much weaker than the lower limits of $10^{-29}G$ is obtained of the order of $10^{-44}G$, certainly much weaker to seed a galactic dynamo or any sort of dynamo amplification. It is also shown here that the strong photon-torsion couplings are well within the COBE CMBR limits. The main ingredient which differs our reasoning from de Sabbata and Sivaram approach is the consist use of dynamo theory and dynamo equation in the universe. One of the main conclusions of the paper is that a stronger massive photon-torsion mass limit is consistent with physics we know. Future prospects include to test these strong couplings in other torsion theories of gravity or at least other examples. It would be also interesting to generalizes this work to include inflation but actually this does not help much dynamo amplification since rapid expansion of the universe tends to make magnetic fields decay faster than here.
\section{Acknowledgements}
 I would like to express my gratitude to A Kostelecky, D Sokoloff, J Beltram-Jimenez and A Maroto for
helpful discussions on the subject of this paper. Financial support from CNPq. and University of State of Rio de Janeiro (UERJ) are grateful acknowledged. Financial support from University of Murcia in Spain is grateful acknowledged.

\end{document}